\documentclass[preprint,prb, amsmath,amssymb,aps,superscriptaddress, nobibnotes]{revtex4-1}
\usepackage{CJK}
\usepackage{graphicx}
\usepackage{color}
\usepackage{float}
\usepackage{xr-hyper}

\makeatletter
\newcommand*{\addFileDependency}[1]{
 \typeout{(#1)}
 \@addtofilelist{#1}
 \IfFileExists{#1}{}{\typeout{No file #1.}}
}
\makeatother

\newcommand*{\myexternaldocument}[1]{%
  \externaldocument{#1}%
  \addFileDependency{#1.tex}%
  \addFileDependency{#1.aux}%
}

\myexternaldocument{suppl}

\usepackage{dcolumn} 
\usepackage{bm}   
\usepackage{amsfonts}
\usepackage[bookmarks=false,colorlinks,citecolor=red]{hyperref}
\usepackage{amsmath}

\usepackage[version=4]{mhchem}
\usepackage{siunitx}

\newcommand{\bfl}{\begin{flushleft}}
\newcommand{\efl}{\end{flushleft}}

\begin{document}

\title{
Antiferromagnetic metal phase in an electron-doped rare-earth nickelate
}

\date{\today}
\author{Qi Song}\thanks{These authors contributed equally: Qi Song, Spencer Doyle.}
\affiliation{Department of Physics, Harvard University, Cambridge, MA, USA}
\author{Spencer Doyle}\thanks{These authors contributed equally: Qi Song, Spencer Doyle.}
\affiliation{Department of Physics, Harvard University, Cambridge, MA, USA}
\author{Grace A. Pan}
\affiliation{Department of Physics, Harvard University, Cambridge, MA, USA}
\author{Ismail El Baggari}
\affiliation{The Rowland Institute at Harvard, Harvard University, Cambridge, MA, USA}
\author{Dan Ferenc Segedin}
\affiliation{Department of Physics, Harvard University, Cambridge, MA, USA}
\author{Denisse C\'{o}rdova Carrizales}
\affiliation{Department of Physics, Harvard University, Cambridge, MA, USA}
\author{Johanna Nordlander}
\affiliation{Department of Physics, Harvard University, Cambridge, MA, USA}
\author{Christian Tzschaschel}
\affiliation{Department of Chemistry and Chemical Biology, Harvard University, Cambridge, MA, USA}
\author{James R. Ehrets}
\affiliation{Department of Physics, Harvard University, Cambridge, MA, USA}
\author{Zubia Hasan}
\affiliation{Department of Physics, Harvard University, Cambridge, MA, USA}
\author{Hesham El-Sherif}
\affiliation{The Rowland Institute at Harvard, Harvard University, Cambridge, MA, USA}
\author{Jyoti Krishna}
\affiliation{Department of Physics, Arizona State University, Tempe, AZ, USA}
\author{Chase Hanson}
\affiliation{Department of Physics, Arizona State University, Tempe, AZ, USA}
\author{Harrison LaBollita}
\affiliation{Department of Physics, Arizona State University, Tempe, AZ, USA}
\author{Aaron Bostwick}
\affiliation{Advanced Light Source, Lawrence Berkeley National Laboratory, Berkeley, CA, USA}
\author{Chris Jozwiak}
\affiliation{Advanced Light Source, Lawrence Berkeley National Laboratory, Berkeley, CA, USA}
\author{Eli Rotenberg}
\affiliation{Advanced Light Source, Lawrence Berkeley National Laboratory, Berkeley, CA, USA}
\author{Su-Yang Xu}
\affiliation{Department of Chemistry and Chemical Biology, Harvard University, Cambridge, MA, USA}
\author{Alessandra Lanzara}
\affiliation{Department of Physics, University of California at Berkeley, Berkeley, CA, USA}
\affiliation{Materials Sciences Division, Lawrence Berkeley National Laboratory, Berkeley, CA, USA}
\author{Alpha T. N’Diaye}
\affiliation{Advanced Light Source, Lawrence Berkeley National Laboratory, Berkeley, CA, USA}
\author{Colin A. Heikes}
\affiliation{NIST Center for Neutron Research, National Institute of Standards and Technology, Gaithersburg, MD, USA}
\affiliation{Joint Quantum Institute, National Institute of Standards and Technology and the University of Maryland, Gaithersburg, MD, USA}
\author{Yaohua Liu}
\affiliation{Neutron Scattering Division, Oak Ridge National Lab, Oak Ridge, TN, USA}
\affiliation{Second Target Station, Oak Ridge National Laboratory, Oak Ridge, TN 37831, USA}
\author{Hanjong Paik}
\affiliation{Platform for the Accelerated Realization, Analysis and Discovery of Interface Materials (PARADIM), Cornell University, Ithaca, NY, USA}
\affiliation{School of Electrical and Computer Engineering, University of Oklahoma, Norman, OK, USA}
\affiliation{Center for Quantum Research and Technology (CQRT), University of Oklahoma, Norman, OK, USA}
\author{Charles M. Brooks}
\affiliation{Department of Physics, Harvard University, Cambridge, MA, USA}
\author{Bet\"{u}l Pamuk}
\affiliation{Platform for the Accelerated Realization, Analysis and Discovery of Interface Materials (PARADIM), Cornell University, Ithaca, NY, USA}
\author{John T. Heron}
\affiliation{Department of Materials Science and Engineering,
University of Michigan, Ann Arbor, MI, USA}
\author{Padraic Shafer}
\affiliation{Advanced Light Source, Lawrence Berkeley National Laboratory, Berkeley, CA, USA}
\author{William D. Ratcliff}
\affiliation{NIST Center for Neutron Research, National Institute of Standards and Technology, Gaithersburg, MD, USA}
\affiliation{Department of Materials Science and Engineering, University of Maryland, College Park, MD, USA}
\author{Antia S. Botana}
\affiliation{Department of Physics, Arizona State University, Tempe, AZ, USA}
\author{Luca Moreschini}
\email{lmoreschini@berkeley.edu}
\affiliation{Platform for the Accelerated Realization, Analysis and Discovery of Interface Materials (PARADIM), Cornell University, Ithaca, NY, USA}
\affiliation{Department of Physics, University of California at Berkeley, Berkeley, CA, USA}
\affiliation{Materials Sciences Division, Lawrence Berkeley National Laboratory, Berkeley, CA, USA}
\author{Julia A. Mundy}
\email{mundy@fas.harvard.edu}
\affiliation{Department of Physics, Harvard University, Cambridge, MA, USA}

\maketitle

\textbf{Long viewed as passive elements, antiferromagnetic materials have emerged as promising candidates for spintronic devices due to their insensitivity to external fields and potential for high-speed switching. Recent work exploiting spin and orbital effects has identified ways to electrically control and probe the spins in metallic antiferromagnets, especially in noncollinear or noncentrosymmetric spin structures. The rare earth nickelate NdNiO$_3$ is known to be a noncollinear antiferromagnet where the onset of antiferromagnetic ordering is concomitant with a transition to an insulating state. Here, we find that for low electron doping, the magnetic order on the nickel site is preserved while electronically a new metallic phase is induced. We show that this metallic phase has a Fermi surface that is mostly gapped by an electronic reconstruction driven by the bond disproportionation. Furthermore, we demonstrate the ability to write to and read from the spin structure via a large zero-field planar Hall effect. Our results expand the already rich phase diagram of the rare-earth nickelates and may enable spintronics applications in this family of correlated oxides.}

In recent years, antiferromagnetic materials have become leading contenders for spintronic devices. In comparison with their ferromagnetic counterparts, more subtle methods to control the ordered spins (writing) and to probe such control (reading) are required \cite{Baltz2018}. The most widely used method in antiferromagnets is the same as that in ferromagnets -- measuring a component of the anisotropic magnetoresistance (AMR) signal\cite{McGuire1975, Daughton1992}. This signal strength is typically limited to just a couple percent of the sample's resistance, however, which limits high-density applications\cite{Jungwirth2016}. This has driven parallel attempts to discover additional read-out methods as well as to identify metallic antiferromagnets with larger AMR signals\cite{Surgers2017, Arpaci2021}. 

\begin{figure*}
\centering
\includegraphics[width =0.88\columnwidth]{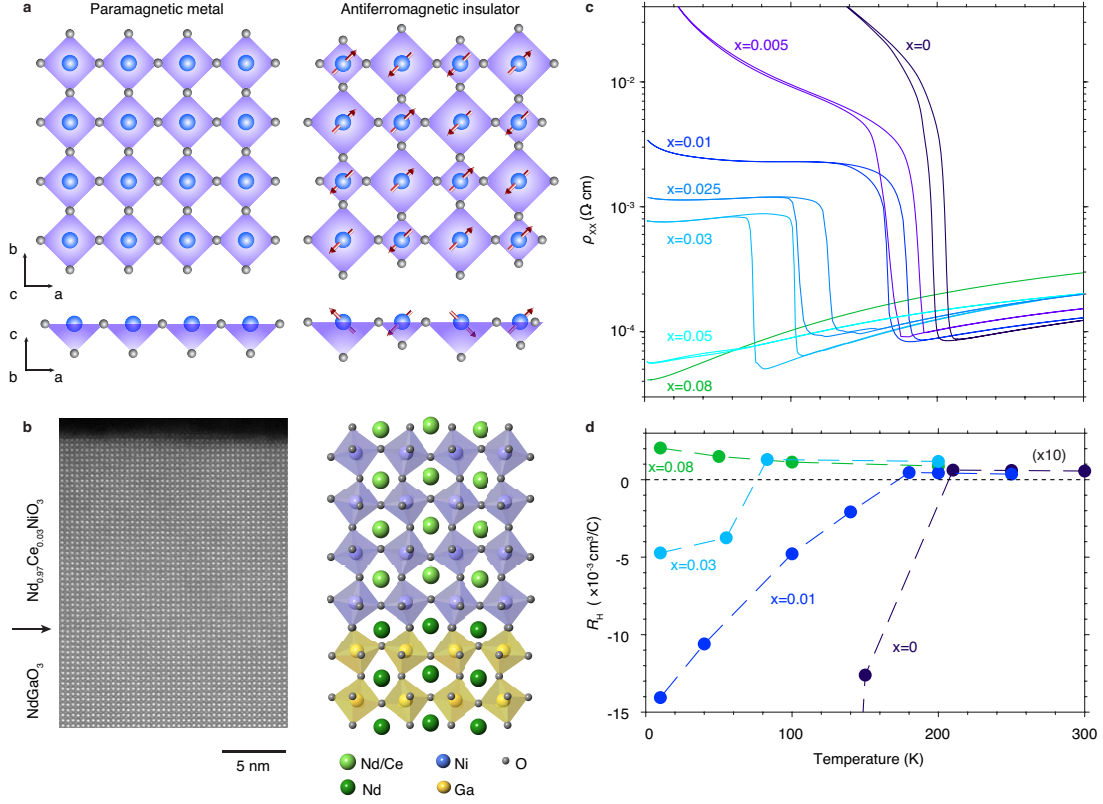}
\caption{\textbf{Structural and electrical characterization of Ce$_x$Nd$_{1-x}$NiO$_3$ thin films.} \textbf{a}, Crystal structure of NdNiO$_3$ viewed in the $ab$- and $ac$-plane. The simplified $(\frac{1}{2},\frac{1}{2},\frac{1}{2})$ bond disproportionation and the $(\frac{1}{4},\frac{1}{4},\frac{1}{4})$ non-collinear antiferromagnetic order below $T_{MIT}$ are illustrated on the right side. \textbf{b}, HAADF-STEM image of a Ce$_{0.03}$Nd$_{0.97}$NiO$_3$ film. The black arrow denotes the coherent substrate/film interface. \textbf{c}, Resistivity $\rho$ \textit{vs.} temperature for Ce$_x$Nd$_{1-x}$NiO$_3$ films synthesized on NdGaO$_3$. \textbf{d}, Hall coefficient $R_H$ \textit{vs.} temperature for the same films, showing a change of sign lowering the temperature across $T_{MIT}$ and $T_{MMT}$. The positive axis is scaled by a factor of 10 for visibility.}
\label{fig:characterization}
\end{figure*}

Here we develop a metallic antiferromagnet and demonstrate nearly an order of magnitude stronger AMR signal over conventional antiferromagnets. We begin with the rare-earth nickelates ($R$NiO$_3$), transition metal oxides which sit at the boundary between localized and itinerant electron behaviors. With the exception of LaNiO$_3$, an unconventional metal\cite{zhou2014mass} with the largest rare-earth cation $R$, the $R$NiO$_3$ compounds undergo a metal-insulator transition (MIT) and a transition from a paramagnetic to a noncollinear antiferromagnetic state\cite{Torrance1992, Scagnoli2006, Bodenthin2011}. In PrNiO$_3$ and NdNiO$_3$ bulk crystals, the onsets of the antiferromagnetic and insulating states are coincident ($T_{MIT}=T_N$)\cite{Medarde1997, Bodenthin2011, Catalano2018, Middey2016}. In addition, at the MIT a breathing-mode distortion occurs where half of NiO$_6$ octahedra are expanded and half are compressed in an alternating pattern: a charge order sets in without any 3$d^6$/3$d^8$ charge fluctuation on the nickel sites, known as ``bond disproportionation''\cite{Johnston2014,Green2016,Varignon2017, chargeorder} (shown schematically in Fig. \ref{fig:characterization}a).

While the rare earth nickelates thus display coupled spin, charge and lattice degrees of freedom, as well as robust metallic and antiferromagnetic states, to date an antiferromagnetic metal phase has not yet been identified\cite{Wang2018}. We choose NdNiO$_3$, which displays a direct transition from a paramagnetic metal to a noncollinear antiferromagnetic insulator at the highest temperature within the $R$NiO$_3$ series \cite{Lacorre1991, GarciaMunoz1992, Frano2013}. Previous studies have found that strong compressive strain\cite{Liu2010, Mikheev2015, Hauser2015} or hole doping on the neodymium site\cite{Cheong1994,Alonso1995,GarciaMunoz1995} tends to suppress the MIT and leaves a paramagnetic metal state, reminiscent of that found in LaNiO$_3$. In contrast, electron doping of bulk crystals through Th$^{4+}$ (ref.~\onlinecite{Alonso1995}) or Ce$^{4+}$ (ref.~\onlinecite{Cheong1994,GarciaMunoz1995}) substitution in the place of Nd$^{3+}$ has been reported to result only in a modest suppression in the MIT transition temperature. Other studies have found electron-doping through oxygen vacancies stabilizes a persistent antiferromagnetic state yet this state is even more resistive than the parent NdNiO$_3$ phase\cite{Li2021,Heo2017}.

Here, we employ electron doping of epitaxial NdNiO$_3$ thin films through substitution of Ce$^{4+}$ on the Nd$^{3+}$ lattice site. We find that cerium substitution rapidly induces a metal-metal transition (MMT) where the low-temperature metallic phase retains the antiferromagnetism on the nickel sites of the parent NdNiO$_3$. In contrast to the metallic phase above MMT and MIT, the Fermi surface below the MMT is largely gapped with spectral weight confined to discrete high-intensity spots. Moreover, in this low-temperature phase we demonstrate heat-assisted magnetic recording with a maximum measured amplitude of 18\%, bringing a family of materials into the spintronics community and suggesting a pathway to stronger AMR readout signals in antiferromagnets.

Epitaxial films of Ce$_x$Nd$_{1-x}$NiO$_3$ were synthesized by reactive oxide molecular-beam epitaxy (MBE) on (100)$_{pc}$ NdGaO$_3$ ((110) in orthorhombic notation for the P\textit{nma} space group), which provides ~1\% tensile strain (see Methods). All crystallographic directions referenced throughout this work follow the pseudocubic (pc) notation. Fig.~\ref{fig:characterization}b shows the high-angle annular dark field scanning transmission electron microscopy (HAADF-STEM) image of the Ce$_{0.03}$Nd$_{0.97}$NiO$_3$ compound. Additional HAADF-STEM and electron energy loss spectroscopy (EELS) maps found in the supplement support the high structural quality and absence of cerium phase segregation or local lattice deformations in these samples (Supplementary Fig.~S1
, Fig.~S5
). X-ray absorption spectroscopy data on the Ce-$M_{4,5}$, Nd-$M_{4,5}$ and O-$K$ pre-peak edges confirm the prototypical Ce$^{4+}$ and Nd$^{3+}$ valence states, and indicate electron doping onto the nickel sites (Supplementary Fig.~S6
).

As seen in Fig.~\ref{fig:characterization}c, cerium substitution results in qualitative changes in the transport properties of the films. The parent compound NdNiO$_3$ shows a MIT at $\sim$200~K with a resistivity increase of about five orders of magnitude \cite{Granados1993}. For intermediate cerium doping (0.01 $\leq$ $x$ $<$ 0.06) the resistivity increases when cooling through the MMT but recovers a $d\rho/dT>0$ behavior, indicative of a metallic phase and a MMT as opposed to the MIT of the parent compound. These samples also demonstrate slight resistive upturns at sufficiently low temperature (below about 10 K), but we believe this to be due to weak localization effects, as suggested by the negative magnetoresistance shown in Supplementary Fig.~S21. 
For higher doping ($x$ $>$ 0.06), the transition is fully suppressed and the films remain metallic at all temperatures. We note the emergence of this MMT transition and fully metallic samples was not observed in previous cerium-doped NdNiO$_3$ powder samples\cite{Cheong1994,GarciaMunoz1995}. Our DFT calculations shown in Supplementary Fig.~S13
 independently demonstrate this behavior as well. In NdNiO$_3$, the sign of the Hall coefficient $R_H$ (Fig.~\ref{fig:characterization}d) changes from positive to negative as the system is cooled below $T_{MIT}$\cite{Hauser2013,Hauser2015}. The same sign change is observed for 0.01 $\leq$ $x$ $<$ 0.06 at $T_{MMT}$, indicating a change in the dominant charge carries from holes to electrons still occurs across the MMT.

\begin{figure*}
\centering
\includegraphics[width = 0.9\columnwidth]{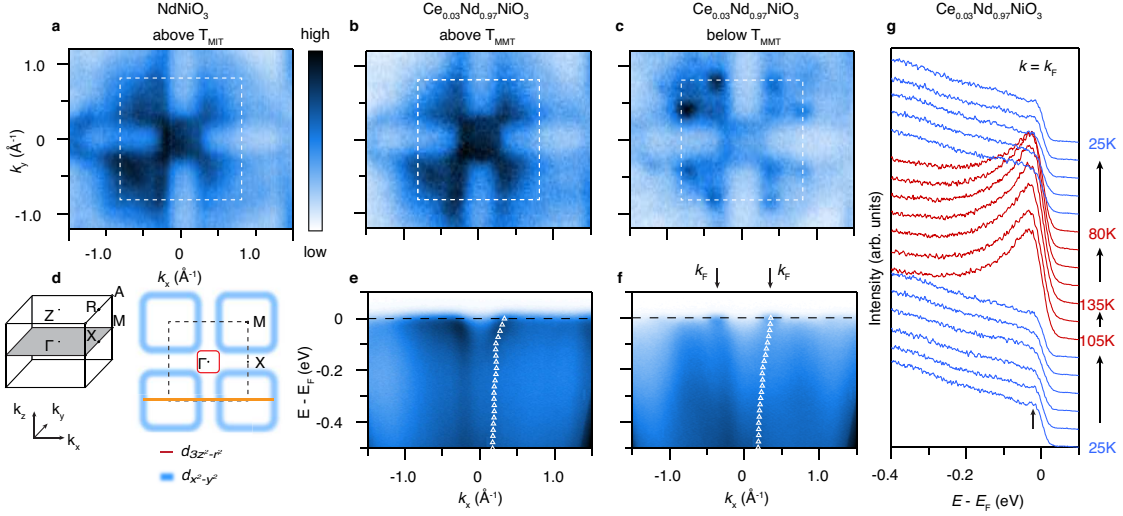}
\caption{\textbf{The electronic transition in Ce$_x$Nd$_{1-x}$NiO$_3$ viewed by ARPES.} \textbf{a,b}, Fermi surfaces of NdNiO$_3$ and Ce$_{0.03}$Nd$_{0.97}$NiO$_3$ measured at 200 K with $h\nu=156$~eV, corresponding to $\Gamma_{004}$. This is above the MIT for NdNiO$_3$ and MMT for Ce$_{0.03}$Nd$_{0.97}$NiO$_3$. The photoemission intensity is coded by the colour bar.  \textbf{c}, same as \textbf{b} but measured at 25 K (below the MMT). \textbf{d}, Sketch of the NdNiO$_3$ three dimensional Brillouin zone, using labels for the tetragonal geometry and Fermi surface on the $\Gamma$XM plane (For NdNiO$_3$, $\overline{\Gamma\mathrm{X}}\simeq0.81\mathrm\AA^{-1}$). The $d_{x^2-y^2}$ contour is blurred to indicate that it is only visible in this plane due to the $k_z$ broadening in the measurement. The horizontal line marks the M-X-M cuts shown in \textbf{e,f}, and corresponding to the maps in \textbf{b,c}, respectively; the markers on the right half of the panels indicate the peak positions extracted from the momentum distribution curves (MDCs) at intervals of 20 meV. The arrows in \textbf{f} point to the Fermi vector $k_F$. \textbf{g}, Energy distribution curves (EDCs) measured at $k_F$ of Ce$_{0.03}$Nd$_{0.97}$NiO$_3$ cycling T from $\sim$25 K (bottom) to $\sim$130 K and then cooling back down to $\sim$25 K (top). The blue/red curves are measured below/above $T_{MMT}$, the arrow indicates the quasiparticle peak at low temperature.}
\label{fig:ARPES}
\end{figure*}

We use angle-resolved photoemission spectroscopy (ARPES) to show that the MMT results in a distinct fermiology at low temperature. Figure \ref{fig:ARPES}a-c show ARPES data measured on the $\Gamma$XM high-symmetry plane of the three-dimensional Brillouin zone sketched in Fig.~\ref{fig:ARPES}d. The Fermi surface of NdNiO$_3$ above $T_{MIT}$ consists of a small electron pocket centered at $\Gamma$ (predominantly of $d_{z^2}$ orbital character) and a larger hole pocket centered at $M$ (predominately of $d_{x^2-y^2}$ character), visible due to the $k_z$ broadening intrinsic to the measurement. This band structure is consistent with prior measurements of NdNiO$_3$ above $T_{MIT}$\cite{Dhaka2015} and is equivalent to that of the fully metallic LaNiO$_3$\cite{Yoo2015, King2014}. As shown in Fig.~\ref{fig:ARPES}b there are no perceptible differences between the Fermi surfaces of Ce$_{0.03}$Nd$_{0.97}$NiO$_3$ and NdNiO$_3$ at 200~K, within the experimental resolution and data quality. In contrast, below $T_{MMT}$ the Fermi surface of Ce$_{0.03}$Nd$_{0.97}$NiO$_3$ does not vanish (Fig.~\ref{fig:ARPES}c) as it does for insulating NdNiO$_3$\cite{Dhaka2015}(see Supplementary Fig.~S7
), but instead turns into a distinctive pattern of discontinuous high-intensity spots located roughly along the original hole pocket contour measured above the transition. This strong reduction of spectral weight of the hole-like Fermi surface is consistent with the change of sign, from positive to negative, observed in $R_H$ at $T_{MMT}$ in Fig.~\ref{fig:characterization}d. For the fully metallic Ce$_{0.1}$Nd$_{0.9}$NiO$_3$, the band structure instead does not show any temperature variation or any noticeable difference with respect to the metallic phase of NdNiO$_3$ (Supplementary Fig.~S11
). The existence of a unique low temperature Fermi surface in the Ce-doped samples hosting a metal-metal transition helps to dismiss concerns of phase coexistence as has been seen in prior nickelate research \cite{Granados1993, Obradors1993}.

In Fig.~\ref{fig:ARPES}e,f we compare the dispersion above and below $T_{MMT}$ for Ce$_{0.03}$Nd$_{0.97}$NiO$_3$ along the MXM line that cuts through one of the high-intensity spots. At high temperature, the band shows a metallic Fermi edge over a broad momentum range, consistent with the $\Gamma$XM plane being close to the boundary of the hole pocket in the $k_z$ direction. At low temperature, the spectral weight at the Fermi level ($E_F$) is suppressed for most of the range, but at some $k_x$ values the metallic edge persists (a more detailed analysis is shown in Supplementary Fig.~S8, Fig.~S10
). Varying the temperature within either metallic phase does not result in any appreciable modification to the maps of Fig.~\ref{fig:ARPES}e,f, while the change is instead abrupt at the transition. Fig.~\ref{fig:ARPES}g shows the evolution of the energy dispersion curves (EDCs) taken at the Fermi vector ($k_F$) determined in Fig.~\ref{fig:ARPES}f, as the temperature is cycled through the MMT . During the warm-up the intensity at $E_F$ suddenly converts into a broad peak at $\sim$105~K that remains unchanged until cooling down to $\sim$80~K, in a hysteresis cycle which perfectly matches that observed in the resistivity curves of Fig.~\ref{fig:characterization}c. Notice that the EDCs of these hot spots retain a metallic edge over the whole temperature range. At the lowest temperature ($\sim$25~K) there is a distinguishable peak-dip lineshape with a small quasiparticle, as opposed to the rest of the Fermi surface which is gapped or pseudogapped below $T_{MMT}$ (see also the analysis in Fig.~S7, S8, S9
). Thus both our electrical transport and ARPES measurements independently show the new metallic phase below MMT in our intermediate cerium doped samples.

The Fermi surface of Fig.~\ref{fig:ARPES}c clearly shows the presence of a new, smaller periodicity in momentum space. Even though it is difficult to sketch the band backfolding due to the presence of multiple bands and of the spectral weight added by the $k_z$ broadening, the pattern formed by the high intensity spots is consistent with a $(\frac{1}{2},\frac{1}{2},\frac{1}{2})$ superstructure, where the $\Gamma$ and A points are brought into each other, as well as Z and M (Supplementary Fig.~S11
). On the ZRA plane weak signatures of this superstructure are present in (metallic) LaNiO$_3$ \cite{Yoo2015}, as well as here in Ce$_x$Nd$_{1-x}$NiO$_3$ for all doping levels above the MIT/MMT (Supplementary Fig.~S11
), which may originate from dynamic charge-order fluctuations as discussed in Supplementary material. \cite{Piamonteze2005,Green2016}. 

\begin{figure}
 \centering
 \includegraphics[width=0.45\columnwidth]{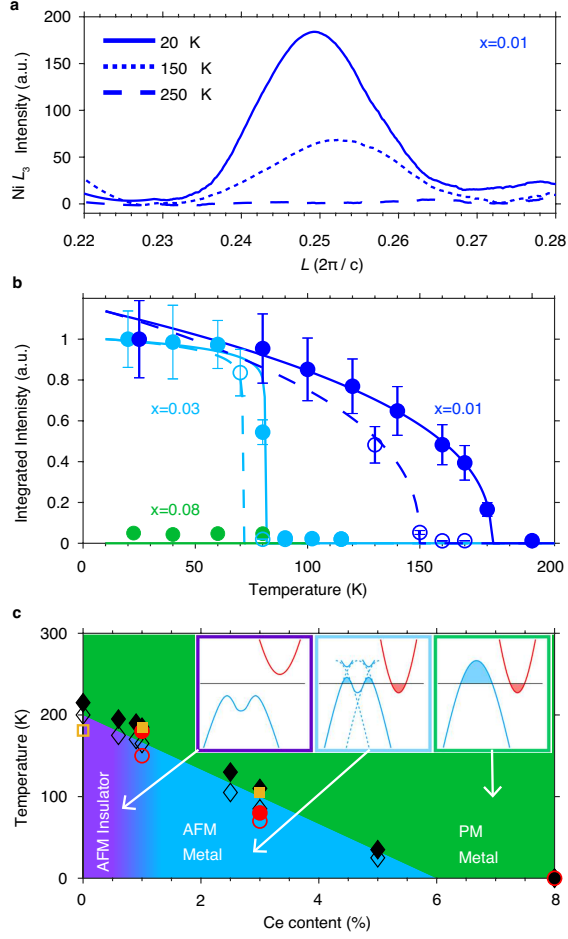}
 \caption{\textbf{Resonant x-ray scattering and phase diagram of Ce$_x$Nd$_{1-x}$NiO$_3$.} \textbf{a}, Ni-$L_3$ edge RXS scans along $L=q_z$ taken above and below $T_{MMT}$ for a Ce$_{0.01}$Nd$_{0.99}$NiO$_3$ sample. We observe the onset of a $q=(\frac{1}{4}, \frac{1}{4}, \frac{1}{4})$ order below 200 K. \textbf{b}, Integrated intensity from the Ni-$L_3$ edge RXS versus temperature for the Ce$_{0.01}$Nd$_{0.99}$NiO$_3$ and Ce$_{0.03}$Nd$_{0.99}$NiO$_3$ samples. The solid and dashed lines are guides to the eye for the warming and cooling data points, respectively. Data are presented as mean values +/- SEM from the n=66 or n=72 pixels located in the region of interest around the scattering vector. \textbf{c}, Electronic and magnetic phase diagram extracted from this study on films synthesized on NdGaO$_3$ substrates. AFM and PM refer to antiferromagnetic and paramagnetic, respectively. The filled and empty symbols indicate the temperatures extracted for warming and cooling cycles, respectively, from electronic transport ($\diamond$), RXS ($\circ$) and ARPES ($\square$). The insets show schematics of the band structure for the three phases. The backfolding in the antiferromagnetic metallic phases is only shown for the hole pocket and without accounting for hybridization for simplicity.}
 \label{fig:RXS}
\end{figure}

To probe the antiferromagnetic order in this low-temperature metallic phase, we use neutron diffraction and site-specific resonant x-ray scattering (RXS). Neutron diffraction (Supplementary Fig.~S15
) shows that the parent NdNiO$_3$ is antiferromagnetically ordered as reported in the literature\cite{Scagnoli2006, Scagnoli2008}. Figure \ref{fig:RXS}a shows RXS data from the Ni-$L_3$ edge of Ce$_{0.01}$Nd$_{0.99}$NiO$_3$, measured along $l$ ($q_z$) with fixed ($h$, $k$) = ($\frac{1}{4}$, $\frac{1}{4}$). We observe an onset of $(\frac{1}{4}, \frac{1}{4}, \frac{1}{4})$ antiferromagnetic order below $T_{MMT}$. The temperature dependence of the (normalized) integrated signal is shown in Fig.~\ref{fig:RXS}b for $x=0.01$, $0.03$ and $0.08$. The absolute integrated signal intensity decreases from the x=0.01 to the x=0.03 sample; see Supplemental figure S15 
and the accompanying discussion for more details. As an additional verification of antiferromagnetic order in the x=0.03 sample, we performed optical second-harmonic generation measurements, which concur with the findings from the RXS measurements presented here (see Supplemental Fig. S18). In contrast to NdNiO$_3$, the magnetic order on the neodymium sites is not present below $T_{MMT}$. This is apparent both in the Nd-$M_{4,5}$ resonant x-ray scattering as well as the absence of observed order in the neutron scattering which is primarily sensitive to the neodymium order (Supplementary Fig.~S15
. Finally, the metallic Ce$_{0.08}$Nd$_{0.92}$NiO$_3$ film lacks observable magnetic ordering on either cation site. Note that we are unable to distinguish a collinear versus noncollinear microscopic ordering from the presented data, however based off of prior works the noncollinear order is the most likely origin of the measured $(\frac{1}{4}, \frac{1}{4}, \frac{1}{4})$ periodicity in our samples\cite{Hepting2018,Scagnoli2006}.

The Ce$_x$Nd$_{1-x}$NiO$_3$ system thus shows three distinct regions as shown in Fig.~\ref{fig:RXS}c. For $x$ $\lesssim$ $0.01$, there is a transition from a paramagnetic metal to an insulator with antiferromagnetic order on both the neodymium and nickel sites. For $0.01$ $\lesssim$ $x$ $\lesssim$ $0.06$, the MIT is replaced by a MMT with antiferromagnetic order remaining on the nickel sites in the low-temperature metallic phase. For $x$ $\gtrsim$ $0.06$, the material lacks evidence of magnetic order, and is likely a paramagnetic metal at all temperatures, similar to LaNiO$_3$\cite{Sreedhar1992} or NdNiO$_3$ under compressive strain \cite{Liu2010, Mikheev2015, Hauser2015}. 

While the insulating phase in the $R$NiO$_3$ family is believed to result from a combination of the lattice distortion and of electronic correlations \cite{Mazin2007,Johnston2014,Green2016,Georgescu2019}, the MMT offers the possibility to partially decouple the two contributions to the transition: Peierls-type and Mott-type\cite{Ruppen2015}. From our data in Fig.~\ref{fig:ARPES}b,c, we argue that in this low-temperature metallic phase the Peierls component of the MIT has taken place, but the additional electrons provided by the Ce$^{4+}$ atoms likely screen correlations just enough that the electron and hole bands are not pushed apart. In this respect, note also that no sizeable bandwidth reduction can be seen in Fig.~\ref{fig:ARPES}e,f, where both above and below $T_{MMT}$ the hole-like band disperses all the way to the onset of the $t_{2g}$ manifold. In this scenario, the charge order is a necessary\cite{Zhang2016b} but not sufficient condition to induce an insulating phase in the $R$NiO$_3$ family. With suppressed or reduced electronic correlations the system remains metallic while retaining the characteristic noncollinear antiferromagnetic order.

\begin{figure}
 \centering
 \includegraphics[width =0.4\columnwidth]{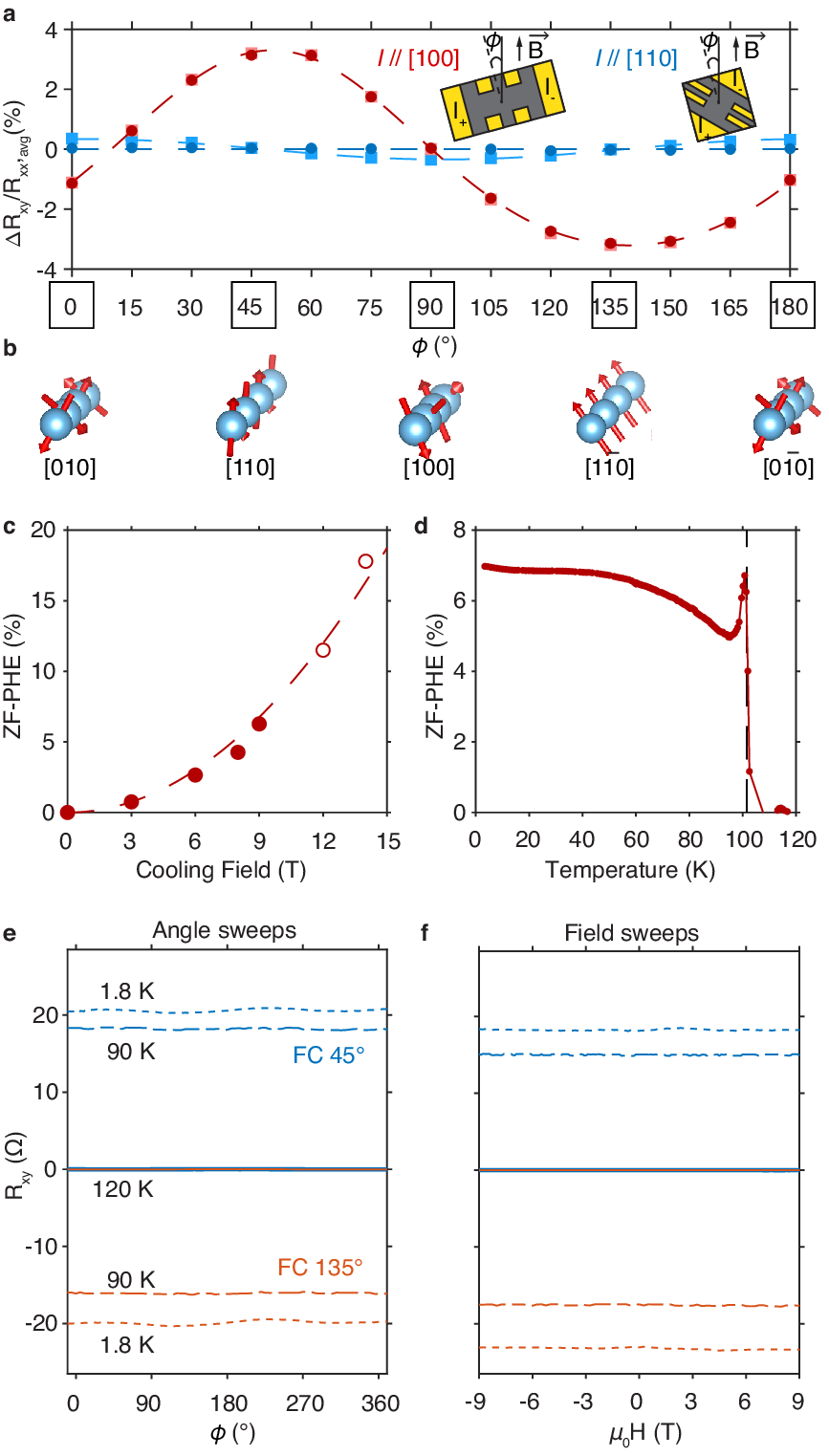}
 \caption{\textbf{Large zero-field planar Hall effect in a metallic antiferromagnetic nickelate. a}, The planar Hall effect magnitude versus in-plane cooling field angle taken at 1.8 K for $I//[100]$ (red) and $I//[110]$ (blue). All data in this figure was taken on samples with $x$=0.02-0.03. Light square points were taken with the field on (9 T) while dark round points were taken after removal of the field (0 T). The dashed lines serve as guides to the eye. The cartoon shows the sample geometry for $I // [100]$. \textbf{b}, Views of the parent compound spin structure for the $(\frac{1}{4}, \frac{1}{4}, \frac{1}{4})$ N\'eel vector viewed along different in-plane crystallographic directions corresponding to the (bolded) in-plane rotation angles above. Dependence of the ZF-PHE magnitude on the strength of the cooling field and temperature (9 T cooling field) are shown in \textbf{c} and \textbf{d}, respectively. High field (12 T and 14 T) scans were performed in an out-of-plane geometry (see Supplementary Fig.~S19
 ). The dashed line in \textbf{c} is a quadratic fit to the data. The stability of the ZF-PHE resistance states to in-plane 9 T field rotations and $\pm$9 T field sweeps are shown in \textbf{e} and \textbf{f}, respectively. Rotations were performed after field-cooling to 1.8 K to both the maximum (45$^o$, blue) and minimum (135$^o$, orange) resistances. In field sweeps, the angle was changed to the opposite extreme after field-cooling to 1.8 K (from 45$^o$ to 135$^o$ and vice versa) followed by a field sweep from +9 T to -9 T.}
 \label{fig:planar_hall}
\end{figure}

Finally, we demonstrate the manipulation of the antiferromagnetic order in this newly stabilized antiferromagnetic metal phase with an external magnetic field, and probe the spin structure via electrical transport. We measure the zero-field planar Hall effect (ZF-PHE) of Ce$_{0.03}$Nd$_{0.97}$NiO$_3$ samples below $T_{MMT}$. Antiferromagnetic materials are expected to display an AMR signal in the presence of a magnetic field as the signal is even (not odd) with the local magnetic moments. Zero-field effects, however, are far less common in antiferromagnets and imply a change to the microscopic spin structure\cite{Marti2014, Kriegner2016, Nair2020}, typically preferential domain formation or canted moment orientation. Figure \ref{fig:planar_hall}a shows the magnitude of the change in R$_{xy}$ (normalized by the average R$_{xx}$ value) versus in-plane angle $\phi$ for current ($I$) applied in two directions: parallel to [100] and parallel to [110]. When current is applied along [100], a significant PHE is present. Removing the external magnetic field has minimal effect on the R$_{xy}$ value, as the ZF-PHE (dark data points) is of the same magnitude as the 9 T PHE (light data points). There is therefore a memory effect written by a cooling field and read by the transverse resistance via the ZF-PHE. Conversely, for current parallel to [110], the 9 T PHE amplitude is reduced by almost an order of magnitude and with the removal of the field, entirely absent (see Methods and Supplementary Figure S19
). Figure \ref{fig:planar_hall}b shows the spin structure of the parent compound NdNiO$_3$ schematically (as found in previous works \cite{Frano2013, Hepting2018}) below the corresponding measurement angles in Fig. \ref{fig:planar_hall}a. There is noncollinearity along [100] and [010], whereas along [110] and $[\Bar{1}10]$, the spins are collinear.

The amplitude of the ZF-PHE for a given cooling field $B_{FC}$ is defined as the overall difference between the peak (at $45^o$ in Fig. 4a) and the trough (in Fig. 4a at $135^o$). We find a maximum amplitude of about 18\%, which is roughly three times stronger than typically reported ZF-PHE magnitudes in antiferromagnets \cite{Jungwirth2016} and, to the best of our knowledge, about equal to the maximum ever reported \cite{Nair2020} (see Supplementary Fig.~S19
). The dependence on $B_{FC}$, Fig.~\ref{fig:planar_hall}c, roughly follows a simple quadratic up to our maximum used cooling field of 14 T, implying that with stronger fields even larger amplitudes are attainable. The temperature dependence, Fig. \ref{fig:planar_hall}d, shows a gradual weakening of the effect as the system approaches its transition and a sudden suppression above the MMT (consistent with the Néel transition). Figure \ref{fig:planar_hall}e,f show constant temperature angle and field sweeps taken after field-cooling the sample at 45$^o$ (135$^o$) into its “high” (“low”) R$_{xy}$ state. The field-cooled states are extremely rigid to external perturbations once they are set; field sweeps over $\pm$9 T and rotations in 9 T have negligible impact on the R$_{xy}$ channel. This is true both at the lowest base temperature of 1.8 K, and at temperatures near but below the transition (90 K shown in the figure).

In ferromagnets, there exist multiple contributions to the AMR signal arising from spin-dependent scattering between conduction electrons and localized magnetic moments \cite{Kokado2012}. The degree of polarization of the conduction electrons critically affects the amplitude of the AMR. In collinear antiferromagnets, the current is strictly not spin polarized. But in noncollinear antiferromagnets, spin-polarized currents are generally allowed\cite{Zelezny2017-noncollinear} and thus the nature of how spin-dependent scattering events contribute to the AMR signal is changed. Given our current direction dependence, it is likely the case that a spin polarized current in a noncollinear antiferromagnet either strengthens or allows additional scattering terms which in turn strengthen the ZF-PHE. Follow-up measurements in other noncollinear antiferromagnets would help clarify whether a large amplitude ZF-PHE is a general trend in such systems. Another natural follow-up experiment would be to attempt to electrically switch these devices using current pulses \cite{Wadley2016}.

Here we explored electron doping of a prototypical transition metal oxide, NdNiO$_3$. The parent NdNiO$_3$ phase is characterized by both spin and charge ordering, perched at the boundary between localized and itinerant electron behaviors: a transition to an insulating, antiferromagnetic state happens in tandem with bond disproportionation. We showed that electron doping can uncover a metallic phase that preserves antiferromagnetic order on the nickel sites and is triggered by a Fermi surface reconstruction induced by charge ordering. We interpret the new phase synthesized as one giving the opportunity of partially decoupling the lattice and electronic contributions to the famous metal-insulator transition characteristic of the perovskite nickelates. We also find that the electron-doped samples display a large zero-field planar Hall effect, and thus offer a new (possibly noncollinear) antiferromagnetic metal of relevance to the spintronics community.

\newpage

\bfl {\bf Acknowledgements} \efl

\noindent This work is supported by the STC Center for Integrated Quantum Materials, NSF Grant No. DMR-1231319. Materials growth was supported by the National Science Foundation (Platform for the Accelerated Realization, Analysis, and Discovery of Interface Materials (PARADIM)) under Cooperative Agreement No. DMR-2039380. This work used resources of the Advanced Light Source, which is a DOE Office of Science User Facility under contract no. DE-AC02-05CH11231. The ARPES work was partially funded by the U.S. Department of Energy, Office of Science, Office of Basic Energy Sciences, Materials Sciences and Engineering Division under Contract No. DE-AC02-05-CH11231 (Ultrafast Materials Program KC2203). Electron microscopy was carried out through the use of MIT nano facilities at the Massachusetts Institute of Technology. Additional electron microscopy work was performed at Harvard University’s Center for Nanoscale Systems, a member of the National Nanotechnology Coordinated Infrastructure Network, supported by the NSF under Grant No. 2025158. A portion of this research used resources at the Spallation Neutron Source (SNS), a DOE Office of Science User Facility operated by the Oak Ridge National Laboratory (ORNL), and resources of the SNS Second Target Station Project. ORNL is managed by UT-Battelle LLC for DOE’s Office of Science, the single largest supporter of basic research in the physical sciences in the United States. This manuscript has been authored by UT-Battelle, LLC, under contract DE-AC05-00OR22725 with the US Department of Energy (DOE). The US government retains and the publisher, by accepting the article for publication, acknowledges that the US government retains a nonexclusive, paid-up, irrevocable, worldwide license to publish or reproduce the published form of this manuscript, or allow others to do so, for US government purposes. DOE will provide public access to these results of federally sponsored research in accordance with the DOE Public Access Plan (http://energy.gov/downloads/doe-public-access-plan). Device fabrication work was performed at Harvard University's Center for Nanoscale Systems (CNS), a member of the National Nanotechnology Coordinated Infrastructure Network (NNCI), supported by the National Science Foundation under NSF Grant No. 1541959. Any mention of commercial products within this paper is for information only; it does not imply recommendation or endorsement by NIST. Work at the NCNR was supported by the Department of Commerce. W. R. acknowledges support from the Department of Commerce. S.D. acknowledges support from the NSF Graduate Research Fellowship Grant No. DGE-1745303. G.A.P. acknowledges support from the Paul \& Daisy Soros Fellowship for New Americans and from the NSF Graduate Research Fellowship Grant No. DGE-1745303. H.E. and I.E. were supported by The Rowland Institute at Harvard. J.N. acknowledges support from the Swiss National Science Foundation under Project No. P2EZP2\textunderscore195686. C.T. acknowledges support from the Swiss National Science Foundation under Project No. P2EZP2\textunderscore191801. Work in the S.-Y.X. group was supported through NSF Career (Harvard fund 129522) DMR-2143177. H.L. and A.S.B. acknowledges support from NSF Grant No. DMR-2045826 and the ASU Research Computing Center for HPC resources. J.A.M acknowledges support from the Packard Foundation and Gordon and Betty Moore Foundation’s EPiQS Initiative, Grant GBMF6760.

\bfl {\bf Author Contributions} \efl

\noindent Q.S., G.A.P., C.M.B. and J.A.M. synthesized the thin-films with assistance from H.P. Q.S. and L.M. performed the ARPES measurements with support from A.B., C.J., E.R., D.F.S., Z.H., and A.L. S.D., Q.S. and J.A.M. performed the electrical measurements with assistance from J.T.H. H.E. I.E. characterized the samples with scanning transmission electron microscopy. G.A.P., S.D., J.R.E., D.F.S., Q.S., D.C.C., A.T.N. and P.S. performed x-ray spectroscopy and scattering measurements. C.A.H., Y.L. and W.D.R. performed the neutron diffraction measurements. J.N., C.T. and S.-Y.X. performed the SHG measurements. B.P. and A.S.B. performed the DFT calculations. J.A.M. conceived and guided the study. Q.S., S.D., L.M. and J.A.M. wrote the manuscript with discussion and contributions from all authors.

\bfl {\bf Competing Interests} \efl
\noindent The authors declare no competing interests.

\clearpage
\newpage 


\begin{thebibliography}{10}
\bibitem{chargeorder}As common practice in the literature, throughout the paper we will interchangeably refer to this as bond disproportionation/ordering or charge ordering with the understanding that its origin is in the hybridization between the O 2$p$ and the Ni 3$d$ states.
\end{thebibliography}

\newpage

\bfl {\bf \large Methods} \efl

\bfl {\bf Growth of Nd$_{1-x}$Ce$_x$NiO$_3$ films} \efl
Ce$_x$Nd$_{1-x}$NiO$_3$ thin films were synthesized on (110) NdGaO$_3$ and (100) LaAlO$_3$ substrates by reactive-oxide molecular-beam epitaxy in a Veeco Gen10 system at PARADIM Thin Film growth Facility at Cornell. The substrates were heated to ~690°C, determined by a thermocouple next to the substrate heater. A mixture of 80\% ozone and 20\% oxygen partial pressure of 2 × 10$^{-6}$ Torr was used during deposition (1~Torr = 0.133322~kPA). Neodymium, cerium and nickel were evaporated from elemental sources. The fluxes of neodymium and nickel were ~1 × 10$^{13}$ atoms$\cdot$cm$^2$/sec and the flux of cerium was adjusted according to the desired doping. Following calibration using a quartz crystal microbalance, the three fluxes were first refined by measuring the thickness of Nd$_2$O$_3$ on (111) (ZrO$_2$)$_{0.905}$(Y$_2$O$_3$)$_{0.09}$ substrate (denoted as YSZ), CeO$_2$ on (111) YSZ substrate and NiO on (100) MgO substrate, respectively. The neodymium and nickel fluxes were then adjusted to produce a stoichiometric NdNiO$_3$ film as judged by the out-of-plane lattice parameter measured by X-Ray Diffraction (XRD) and the resistivity. The thickness of deposited films varies between 14 nm and 23 nm. The structure was characterized by a PANalytical Empyrean x-ray diffractometer at 45 kV and 40 mA with Cu K$\alpha_1$ radiation.The X-ray diffraction \(\theta\)-2\(\theta\) scans indicate that all of the Ce$_x$Nd$_{1-x}$NiO$_3$ films are single phase grown on NdGaO$_3$ (Fig.~S2
) and LaAlO$_3$ (Fig.~S3
) substrates. The comparison of phase diagram of the films grown on NdGaO$_3$ and LaAlO$_3$  is shown in Fig.~S4. 

\bfl {\bf Resonant x-ray scattering} \efl
Resonant x-ray scattering measurements were performed on Ni-$L_3$ and Nd-$M_5$ edges at Beamline 4.0.2 at the Advanced Light Source at Lawrence Berkeley National Laboratory. Reciprocal space maps (RSMs) were generated by aligning to the signal peak after performing $L$-rod scans, and capturing fixed geometry CCD images for bins around the peak location. Note that the deviation of the peak maximum from the expected $L=0.25$ is small enough to be due to an alignment offset, and is not directly indicative of any incommensurate order. Reciprocal space maps were measured at a series of temperatures upon warming and cooling each sample and regions of interest were defined around the observed order peaks. To extract integrated intensities, constant size regions of interest were defined around the peak in the RSMs and the total intensity contained was summed. Background intensities (used for background subtraction) were collected in the diffuse regions around the ordering peaks. Small $Q_x$ and $Q_y$ offsets were added relative to the location of the $(\frac{1}{4},\frac{1}{4},\frac{1}{4})$ peak to define a region of background intensity.

\bfl {\bf X-ray absorption spectroscopy} \efl
We use Beamlines 6.3.1 and 4.0.2 at the Advanced Light Source at Lawrence Berkeley National Laboratory to perform x-ray absorption spectroscopy (XAS). Measurements were conducted in the total electron yield geometry. Each spectra shown is the average of 4-16 individual scans. The Nd-M$_{4,5}$ and Ce-M$_{4,5}$ spectra were normalized to the pre-edge intensity. The O-$K$ edge spectra were normalized to the integrated peak intensity.

\bfl {\bf Transport measurements} \efl
Cr(10 nm)/Au(100 nm) electrical contacts were deposited in the pattern shown by the cartoon of Fig. \ref{fig:planar_hall}a with an electron-beam evaporator. Channels were etched in with a diamond scribe. A typical device was roughly 3mm x 5mm. Transport measurements were performed in a Quantum Design Dynacool Physical Property Measurement System (PPMS) for fields up to 9 T. In plane rotations were performed using the Quantum Design horizontal rotator option. Measurements performed at 12 T and 14 T were conducted in a 14 T Quantum Design PPMS. Hall coefficients were calculated from linear fits of antisymmetrized field sweeps measured from $\pm$ 9 T; a representative example is shown in Supplemental Figure~S22
. In Fig. \ref{fig:planar_hall}e/f, a vertical offset due to contact misalignment was removed.

\bfl {\bf Angle-resolved photoemission} \efl
The ARPES data was measured at the beamline MAESTRO of the Advanced Light Source using a Scienta 4000 electron analyzer. The angular resolution is better than 0.1$^\circ$ and the energy resolution varied between 10~meV and 25~meV depending on the setting chosen. The light polarization was set to $p$. In this configuration when the sample is facing the analyzer slit with the $a$ axis in the horizontal ($x$) direction the electric field is even in the $xz$ mirror plane plane containing the surface normal and therefore the measurement is sensitive to the $d_{x^2-y^2}$ and $d_{z^2}$ orbitals along $k_x$, which are both even in the same plane. Prior to the ARPES experiments, the films were annealed in a a partial oxygen pressure of ~1 × 10$^{-5}$ with $\sim$10\% ozone at a temperature of ~420 °C for approximately one hour.

\bfl {\bf Neutron Scattering Measurements} \efl
Elastic neutron scattering data were collected at the SPINS cold triple-axis spectrometer at the NIST Center for Neutron Research (NCNR) and at the CORELLI elastic diffuse scattering spectrometer\cite{ye2018implementation} at the Oak Ridge National Laboratory Spallation Neutron Source (SNS). The SPINS measurements were taken at a fixed neutron energy of 5 meV with a collimation and filter setup of 80’-Be-80’-Be using a standard closed cycle refrigerator for sample temperature control. \par
CORELLI is a quasi-Laue time-of-flight instrument with an incident neutron wavelength band between 0.65 and 2.9 Å. The two-dimensional detector array of CORELLI spans from -19$^{\circ}$ to 147$^{\circ}$ in the horizontal plane and from -27$^{\circ}$ to 29$^{\circ}$ vertically. Therefore, a large three-dimensional volumetric reciprocal space is surveyed for a single sample orientation. The sample was mounted to an Al pin at the bottom of a close cycle refrigerator, which provided a base temperature of 3 K at NCNR and 6 K at CORELLI. The sample was mounted with the (h k 0)$_{o}$ plane horizontal, and the vertical rotation axis was along the (0 0 l)$_{o}$ axis. The experiments were conducted by first rotating the film through 120$^{\circ}$ in 1$^{\circ}$ step to survey features in the large reciprocal space, and then data were collected for an extended time with particular sample orientations optimized for the selected reciprocal regions.

\bfl \textbf{Scanning Transmission Electron Microscopy} \efl
High-angle annular dark-field STEM measurements were performed either on a JEOL ARM 200F (on the Ce 4\% sample) or a Thermo-Fisher Scientific Titan Themis Z G3 (on the Ce 12\% sample) both operating at 200 kV. The convergence angle was either 19.6 or 22 mrad and the collection angle range was ~68 mrad-280 mrad. Fast-acquisition frames were collected, aligned and summed to minimize drift and obtain high signal-to-noise-ratio.
The interfacial strain maps were extracted using the method described in the reference \cite{Goodge2022}.

\bfl {\bf Electron Energy Loss Spectroscopy} \efl
Electron energy loss spectroscopy (EELS) were performed on the Thermo-Fisher Scientific Titan Themis Z G3 equipped with a Gatan Continuum spectrometer and imaging filter. The microscope was operated at 200 kV and 100 pA current. Dual (zero-loss and core-loss) EELS spectrum images were acquired simultaneously to correct for energy drift. After determining a region of interest in a zone axis, the electron beam was blocked for 15 minutes to stabilize and minimize the stage drift. The total spectrum image time was maintained between 30 to 100 seconds to minimize stage drifting. EELS maps were acquired with a 6 Å pixel-size sampling, a 30 ms pixel dwell time and 0.75 eV/channel electron dispersion.

To extract the elemental composition maps, EELS spectra were background subtracted up to the relevant elemental edge and then integrated. The background was modelled as a linear combination of power laws and the background window was selected to be as wide as possible to obtain good fits. The energy windows were as follows: Ni-$L_{2,3}$ edge (851-878 eV), Ce-$M_{4,5}$ edge (881-913 eV), Nd-$M_{4,5}$ edge (975-1010 eV).

\bfl {\bf Computational Methods} \efl
Density functional theory (DFT)-based calculations were performed using the all-electron, full potential code {\sc wien2k} based on the augmented plane wave plus local orbital (APW+lo) basis set \cite{w2k}. For the exchange-correlation functional, we have used the Perdew-Burke-Ernzerhof (PBE) implementation of the generalized gradient approximation (GGA) \cite{pbe}. In the spin-polarized calculations, in order to treat the correlated Ni$-d$ states, we added static mean-field correlations as implemented in the DFT+$U$ framework \cite{sic} with an on-site Coulomb $U$ repulsion ranging from $2$ to $5$ eV and a non-zero Hund's coupling $J = 0.7$ eV to account for the anisotropies of the interaction \cite{pickett_lsda+u}. The results we present are consistent for this range of $U$ values.

Phonon calculations were performed using Vienna Ab-initio Software Package (VASP) \cite{vasp1,vasp2} which implements projector augmented waves formalism of DFT with an interface to Phonopy \cite{phonopy} for plotting the phonon dispersion. The same DFT+$U$ framework as above was added to the GGA/PBE functionals. We employed an electronic energy tolerance of $10^{-8}$ eV and a force tolerance of $10^{-3}$ eV/\AA~ for structural relaxations, with a Gaussian smearing of 0.005 eV, and an electronic momentum $k$-point mesh of $2 \times 2 \times 2$ for a $4 \times 4 \times 4$ supercell structure. We obtain similar phonon instabilities with an electronic momentum $k$-point mesh of $5 \times 5 \times 5$ for a $2 \times 2 \times 2$ supercell structure.

\bfl {\bf Data Availability} \efl
\noindent The data presented in the figures and other findings of this study are available from the corresponding authors upon reasonable request.

\end{document}